# Novel Spacetime Concept and Dimension Curling up Mechanism in Neon Shell

Kunming Xu
*Environmental Science Research Center,*
*Xiamen University, Fujian Province 361005, China*
kunmingx@yanan.xmu.edu.cn

Euclidean geometry does not characterize dynamic electronic orbitals satisfactorily for even a single electron in a hydrogen atom is a formidable mathematical task with Schrödinger's equation. Here the author puts forward a new spacetime concept that regards space and time as two orthogonal, symmetric and complementary quantities. They are inherent physical quantities that cannot be divorced from physical objects themselves. In two-dimensional helium shell, space and time are instantiated by two interactive 1s electrons; in four-dimensional neon shell, space and time dimensions blend into four types of curvilinear vectors represented by 2s, $2p_x$, $2p_y$, and $2p_z$ electronic orbitals. The description of electronic orbitals constitutes an explanation of canonical spacetime properties such as harmonic oscillation, electromagnetism, and wave propagation. Through differential and integral operations, the author formulates a precise wavefunction for every electron in an inert neon atom where spacetime, as dimensional graduated by ten electrons, is continuous, and trigonometric function is the mechanism for dimension curling up. This fresh spacetime view based on dimensional interpretation of complex functions is an extension of classical mechanics and is compatible with relativity and quantum physics. It brings sharp insight into the geometries of 2p-orbitals and has broad support from chemistry.

## 1. Introduction

Since antiquity, humans have believed that they know about space and time because of their direct experience, but scientific conception of them turns out to be elusive. This is not because space and time are so complex but they are so fundamental that there is not any preceding rule for reference. Einstein set forth relativity based on Minkowski coordinates that couple three-dimensional space with one-dimensional time through events, but gained results such as time dilation and space contraction that cannot be easily explained by traditional Euclidean space and Newtonian time. In 1919, Kaluza unified Maxwell's electromagnetism and Einstein's general relativity and gravity by adding a fifth dimension [1]. He suggested that the fifth space dimension normally curls up and is hidden from direct visualization. To date, the detailed mechanism for dimension curling up, however, has never been demonstrated either mathematically or physically. Turning our focus to microcosm, quantum mechanics uses statistical probability and quantum numbers to describes electrons. This uncertainty approach is radically different from that of classical mechanics. Moreover, relativity and quantum physics remain detached and must be unified. Since space and time are the most relevant and underlying physical quantities for relativity, classical and quantum mechanics, the lack of unification among the three fields prompts us to search for new space and time concepts on a more profound level than the homogeneous Euclidean space and uniform Newtonian time. This paper explores the basic concept of spacetime starting from two-dimensional helium shell and further proposes the mechanism for dimensions curling up in four-dimensional neon shell.

## 2. Two-dimensional spacetime

In spite of the great progress made in relativity and quantum mechanics during the last century, our current conception of spacetime is still framed by three-dimensional Euclidean geometry and one-dimensional Newtonian time. To establish a more fundamental spacetime concept other than that framework, we need to be cautious on any presumptions that we have inadvertently introduced. Space and time are the foundation and background of all sciences. Let's discard every antecedent belief and premise except saying that space and time are a pair of the most fundamental physical quantities. They are inherent properties that cannot be divorced from physical entity as if two sides of a coin cannot be divorced from the coin itself.

To expatiate on what space and time are from scratch, we notice that there are two electrons in an inert helium atom. We may associate space with one electron, and time with the other, or in a more general dynamic manner as will be introduced. Helium shell is such a conservative system that both electrons should best represent the two basic dimensions. It is by this approach that we explore the property of space and time through the description of electronic orbitals.

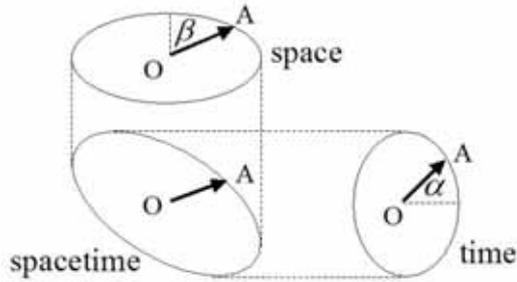

Fig. 1 Time and space components of an electron within helium shell controlled by $\alpha$ and $\beta$ radians, whose values depend on the position of $A$ along an imaginary circular track.

Fig. 1 characterizes the motion of an electron in helium shell as harmonic oscillation with its time component $\Psi$ and space component $\psi$ satisfying:

$$\frac{d^2\Psi}{dt^2} = -\omega^2 \Psi \qquad (1)$$

$$\frac{\partial^2 \psi}{\partial l^2} = -\frac{1}{r^2}\psi \qquad (2)$$

$$v = \omega r \qquad (3)$$

where $t$ and $l$ are time and space dimensions respectively, $\omega$ denotes angular velocity, $r$ is orbital radius, and $v$ is velocity. Bear in mind that we are dealing with two-dimensional spacetime where space and time have non-classical meanings, so do other related quantities. A typical solution to eq. (1) is:

$$\Psi = C_1(\cos\alpha - i\sin\alpha) \qquad (4)$$

$$-\frac{\partial \alpha}{\partial t} = \omega \qquad (5)$$

where $C_1$ is a constant, $\alpha$ is a radian angle, and the negative sign before the derivative implies contrary directions of $\alpha$ angle movement relative to time dimension orientation. Similarly, the wavefunction describing space component of the electron is:

$$\psi = C_2(\cos\beta + j\sin\beta) \tag{6}$$

$$\frac{\partial \beta}{\partial l} = \frac{1}{r} \tag{7}$$

where $C_2$ is a constant, $j$ is a complex number notation like $i$, but it describes imaginary space instead of imaginary time component. Since space and time components are orthogonal, we express electronic wavefunction with the product of both components:

$$\Omega = \Psi\psi \tag{8}$$

$$\Omega = C_1 C_2 (\cos\alpha\cos\beta - i\sin\alpha\cos\beta - ij\sin\alpha\sin\beta + j\cos\alpha\sin\beta) \tag{9}$$

Here multiplication makes sense for two orthogonal quantities. However, to understand the meaning of the wavefunction, we must decipher the meanings of $i$ and $j$ notations. In principle, a complex number is introduced when a real number cannot express a two-dimensional vector. Under Cartesian X-Y coordinates, the identifier $i$ is an operator casting a real number in X-axis into an imaginary component along Y-axis. Logically, the imaginary and the real parts of a complex number belong to different space dimensions.

How to express various spacetime dimensions in two-dimensional helium shell? If we start with a dimensionless quantity, change it in the direction of reducing a time dimension and increasing a space dimension, and reverse the other way around to complete a cycle, then we get four types of dimensional quantities out of the possible one-dimensional space and one-dimensional time combinations as represented by (1, $\omega$, $v$, $r$) with SI units of (1, 1/s, m/s, m) respectively, using meter and second to denote space and time units. Thus, the significance of complex notations in eq. (9) may be interpreted as:

$$\begin{pmatrix} 1 & i \\ ij & j \end{pmatrix} = \begin{pmatrix} 1 & \omega \\ v & r \end{pmatrix} \tag{10}$$

so that each of the four terms on the right-hand side of eq. (9) is a characteristic root or electronic state $\Omega_i$:

$$\begin{pmatrix} \Omega_0 \\ \Omega_1 \\ \Omega_2 \\ \Omega_3 \end{pmatrix} = C_1 C_2 \begin{pmatrix} \cos\alpha\cos\beta \\ -\omega\sin\alpha\cos\beta \\ -v\sin\alpha\sin\beta \\ r\cos\alpha\sin\beta \end{pmatrix} \tag{11}$$

Since there are only two electrons in helium shell, each electronic orbital must include two adjacent roots. For instance, a wavefunction may take the form of ($\Omega_0 + \Omega_1$). Considering $\omega^2 = -1$ and $r^2 = -1$, there are interesting calculus relationships between the four roots:

$$-\frac{\partial \Omega_0}{\partial t} = \Omega_1 \tag{12}$$

$$\int \Omega_1 dl = \Omega_2 \tag{13}$$

$$-\frac{\partial \Omega_2}{\partial t} = \Omega_3 \tag{14}$$

$$\int \Omega_3 dl = \Omega_0 \tag{15}$$

Mathematics expresses physics. Each mathematical relationship has its physical correspondence and reflects electronic behavior in dynamic process. To specify, when an electron is at the state of ($\Omega_0 + \Omega_1$), its $\Omega_0$ component is transforming into $\Omega_1$ component according to eq. (12) while $\Omega_1$ is converting into $\Omega_2$ via eq. (13). In other words, the electron is shifting its state from ($\Omega_0 + \Omega_1$) to ($\Omega_1 + \Omega_2$), losing a time dimension and gaining a space dimension, while another electron in helium shell evolves from state ($\Omega_2 + \Omega_3$) to state ($\Omega_3 + \Omega_0$). Eqs. (12) to (15) form a loop describing the harmonic oscillation of electrons in spacetime. In short, electronic motion follows simultaneous differential and integral operations.

In the simplest sense, oscillation of electrons is similar to a pendulum. However, if electrons were orbiting around the nucleus like planets around the sun kinematically, then they would emit energy due to their high frequency. As a result, the system would be damped quickly. To overcome this, electrons must oscillate through changing states so that both electrons exchange energy, i.e., each electron receives the quantity emitted by another so that helium shell remains conservative. Here the electron is revolving in the sense that it changes physical state continuously and periodically as the state point, $A$, orbits around the origin O (Fig. 1). The circular track of point *A* represents the pathway of electronic state transformation rather a kinematic movement.

Combining eqs. (12) and (13) yields

$$-\frac{\partial \Omega_0}{\partial t} = \frac{\partial \Omega_2}{\partial l} \tag{16}$$

which means that the changing rate of one electron in time is compensated by the varying rate of another in space. This equation is similar to Faraday's law:

$$\nabla \times E = -\frac{\partial B}{\partial t} \tag{17}$$

In two-dimensional system, operator $\nabla \times$, naturally declines into $\partial/\partial l$. If we treat $\Omega_0$ as a magnetic field and $\Omega_2$ as an electric field, then eq. (16) is another expression of Faraday's law. This indicates that electronic oscillation is an electromagnetic action. Furthermore, if we treat $\Omega_0$ as a probability density function and $\Omega_2$ as probability current, then eq. (16) also indicates that a change in the density in region *l* is compensated by a net change in flux into that region. This agrees with quantum mechanics on probability.

As electrons are oscillating between space and time, increasing a space dimension is accompanied by decreasing a time dimension, and vice versa. In other words, expansion of space is undergoing with release of time wrinkles whereas contraction of space results in condensation of time. When space fully unfolds, it loses all density and wraps back according to its cycle, so does time. Space and time components are coupled together in such an intimate way that spacetime is a finite and yet unbounded continuum, which agrees with relativity. Here space is no more a three-dimensional volume, and time is no more a unidirectional flow. Two dimensions refer to two equally and complementarily functional aspects of electrons, i.e. two modes in spacetime. Space still has X, Y, and Z orientations in Euclidean geometry, so does time. But it is more proper to adopt sine and cosine functions in various dimensions (1, $\omega$, $v$, $r$) to characterize their intricacy than to use homogeneous X, Y, and Z coordinates with linear algebra.

From the perspective of waves, if we regard each root, $\Omega_i$, as a waveform, then each electronic orbital is composed of two adjacent roots, and therefore covers two waveforms. Because every pair of adjacent roots are exactly one dimension apart, separated by either a time or a space dimension exactly, the two waveforms are orthogonal. An electronic wave propagates from one waveform to another following eqs. (12) to (15). However, a differential operation on a trigonometric function with respect to a time dimension, such as that dictated by eq. (12), does not physically happen in a flash, but it is carried out gradually and smoothly. When $\cos\alpha$ receives the differentiation order, the angle $\alpha$ is then rotating gradually up to ($\pi/2+\alpha$), at which point the differential operation completes. It goes without saying that the velocity of $\alpha$ rotation, or angular velocity $\omega$, determines the speed of the differential process and hence the period of electronic oscillating cycle. In this way, we have explained electronic motion by calculus and the implementation of calculus by trigonometric functions. Mathematical expressions represent dynamic physical processes.

To summarize, this section started from mere assumption of harmonic oscillation of two electrons in helium shell, and then interpreted complex wavefunctions in terms of possible two dimensional spacetime quantities. Two electrons at a specific moment constitute two basic dimensions, time and space, which are separated by π/2. For example, an electron ($\Omega_0+\Omega_1$) may represent time as $\alpha$ equals 0 while the other electron ($\Omega_2+\Omega_3$) may represent space as $\alpha$ equals π/2. At that specific moment, both $\Omega_1$ and $\Omega_3$ vanish so that one electron, $\Omega_0$, indicates time while the other, $\Omega_2$, indicates space dimension. When 0< $\alpha$ <π/2, both electrons contain a mix of space and time components. This new outlook of spacetime is beyond our usual mental concept, but it is the beginning of a series of discoveries in the following sections.

## 3. Quaternity space

We have initiated a new space and time concept in two-dimensional helium shell, but the most important spacetime is of four dimensions. The best way to delineate a four-dimensional space is to give four geometric shapes corresponding to four space dimensions. To this end, we shall examine the property of a circle and a spherical layer from the standpoint of their centers.

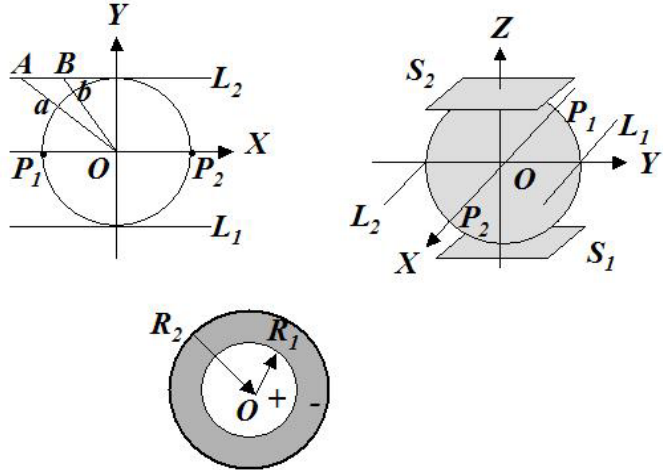

Fig. 2 Orthogonal analyses of space dimensions in a spherical layer. A circle can be projectively decomposed into two lines and two points; a spherical surface into two flat planes $S_1$ and $S_2$, two lines $L_1$ and $L_2$, and two points $P_1$ and $P_2$; and a spherical layer can be represented by subtracting an inner solid sphere $R_1$ from an outer sphere $R_2$.

As shown in Fig. 2, a circle is composed of two lines and two points. For every point $a$ in the circle, we can always find a point $A$ in the lines corresponding to it except at point $P_1$ and $P_2$. Conversely, for every point $B$ in lines $L_1$ or $L_2$, we can always find a point $b$ in the circle corresponding to it. Likewise, there are projective correspondences between a spherical surface and two flat planes, two straight lines, and two points. We shall identify the two points $P_1$ and $P_2$ as a one-dimensional geometric element, the two parallel lines $L_1$ and $L_2$ as a two-dimensional element, and the two flat planes $S_1$ and $S_2$ as a three-dimensional element. These three geometric elements are a distorted expression of the spherical surface. They are mutually exclusive and oriented in X, Y, and Z directions respectively. At X direction, $P_1$ and $P_2$ are two points on the spherical surface; at Y direction, $L_1$ and $L_2$ are two lines corresponding to two semicircular arcs on the spherical surface; and at Z directions, $S_1$ and $S_2$ are two flat plan corresponding to two hemispherical surface on the spherical surface.

Moreover, a solid sphere has one additional dimension in the radial direction compared with a spherical surface, so it contains four geometric elements, the fourth one being a radial vector that is orthogonal to the spherical surface. A spherical layer is an outer sphere minus a concentric inner sphere. Thus, two radii, two planes, two lines, and two points accurately characterize a spherical layer. These four elements constitute four orthogonal geometries. Here orthogonality means that quantities are not only mutually exclusive with π/2 radian angle interval but also of different space dimensions.

As we study a spherical layer, the outer sphere can be treated as an unfolding four-dimensional space whereas the inner sphere is a zero-dimensional point encapsulating all complexities within it. In Euclidean geometry, a point is assumed to be infinitesimal in size, but here we regard a point as possessing the size of the inner sphere, and lines and planes have certain thickness, too. The outer radius, two planes, two lines, two points, and the inner radius are four, three, two, one, and zero-dimensional elements, respectively, because they correspond to the highest dimension of an outer sphere, a spherical surface, a circle, two dislocated points, a concentric inner sphere, respectively (Fig 3).

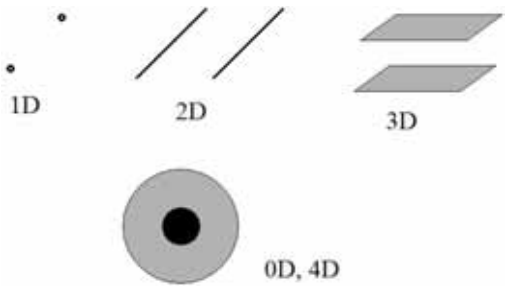

Fig. 3 Four distinct space elements, two points, two lines, two planes, and two spheres for characterizing a spherical layer.

If we start from a four-dimensional outer sphere $E_0$, and reduce a dimension from it, then we get a spherical surface $E_1$. The immediate dimension that was reduced is a radial quantity. Such an action is like taking away 8000 from the number 8888 because a radial vector is a four-dimensional element. We express this process in a differential way (or in a quotient way) as:

$$\frac{dE_0}{dl_0} = E_1 \tag{18}$$

where $l_0$ denotes a four-dimensional radial element, the largest dimension within $E_0$. If we further reduce a dimension from the spherical surface, we get a circle $E_2$. The highest space dimension that was taken away is a three-dimensional element, analogous to 800 in the number 888. Likewise, reducing a dimension from the circle leaves two points $E_3$ dislocated off the center; and further reducing a dimension from $E_3$ results in an inner spherical origin $E_4$. These geometrical transformations can be expressed by differential operations:

$$\frac{dE_1}{2dl_1} = E_2 \; ; \quad \frac{dE_2}{2dl_2} = E_3 \; ; \quad \frac{dE_3}{2dl_3} = E_4 \tag{19}$$

where $2l_1$ is a three-dimensional element (represented by two parallel planes $S_1$ and $S_2$), $2l_2$ is a two-dimensional element (represented by two parallel lines $L_1$ and $L_2$), and $2l_3$ is a one-dimensional element (represented by two symmetric points $P_1$ and $P_2$ off spherical origin). Hence we have defined the convention: A differential operation upon a geometrical shape is always operated with respect to its highest or most outstanding dimensional element, analogous to taking away the leftmost digit from a number.

It is interesting to find that, after reducing four dimensions sequentially, a sphere produces the same shape of different size. The four dimensions reduced are a radius, two planes, two lines, and two points in sequence, and the final result is an inner sphere represented by a smaller radius. An outer sphere reducing four dimensions sequentially results in a smaller sphere. Depending on the scope of concern, the inner sphere may serve as an outer sphere for further decomposition. Conversely, an outer sphere may serve as an inner sphere for forming an even larger spherical layer. Thus a sphere may be treated as either a four-dimensional outer sphere or a zero-dimensional inner sphere. The differential operations from a larger sphere to another small sphere give scalability of the four elements, which can be repeatedly used for differentiating a large multiple-layered sphere.

Four space elements (radii, planes, lines, points in Fig. 3) that we defined are of various spatial significances. This is somewhat analogous to number expression. For example, in the number 8888, the first 8 means eight thousand, the second means eight hundred, and so on and so forth, the four digits actually denote different magnitudes. In fact, it is such a kind of structural grading that makes it feasible to express a large number. Quaternity space follows the same logic. Four space elements form a ladder (i.e. four, three, two, one, and zero) in terms of dimensional magnitude, which is contrary to Cartesian coordinates where dimensional properties of X, Y, and Z directions are assumed to be the same. Due to their scalability and grading, quaternity dimensions make it feasible to express spherical structures of multiple layers. Quaternity space means four distinct space dimensions in a spherical layer. It is evident that the atomic space such as neon atom is spherical rather than linear. So it is our intention to apply this new space concept to electronic orbitals.

## 4. Quaternity spacetime and rotation operation

We are apt to take it for granted that time is one-dimensional and space is three–dimensional. After giving up this presumption and treating space and time as complementary and symmetric quantities, we get four kinds of space and time combinations in a spherical layer (Tab.1). The outer shell of a neon atom is a spherical layer composed of eight electrons that can be expressed spatially by the octet: two spheres, two points, two lines, and two planes.

The number of time dimension is complementary to the number of space dimension. When the total dimensions are fixed for a given electron, the electron reducing a dimension in space inevitably increases a dimension in time accordingly, and vice versa. We define the medium of three-dimensional space with one-dimensional time as a vitor, meaning vitality; we define the medium of two-dimensional space with two-dimensional time as a metor, which connotes metamorphosis in the midway; and define the medium of one-dimensional space with three-dimensional time as a relator, which serves as a tendency reference. These three quantities are all curvilinear vectors, referring to different space and time combinations. We use a scalar to refer to four-dimensional space with zero-dimensional time of outer sphere or zero-dimensional space with four-dimensional time of inner sphere. We may associate an expanding space with a positive scalar, then a contracting space is associated with a negative scalar. Strictly speaking, a scalar in quaternity spacetime is also a curvilinear vector because it is dynamic in nature and has an expanding or contracting direction. Finally we define any of the four quantities, the scalar, the relator, the metor, and the vitor, as a quadrant.

Tab. 1 Quaternity spacetime and wavefunctions based on four orthogonal geometries and their distorted correspondences in a spherical layer where space and time dimensions are complementary.

| Quadrant | Space dim | Time dim | Projective geometries | Spherical layer correspondences | Wave-Functions | Orbitals |
|---|---|---|---|---|---|---|
| Scalars | 0 or 4 | 4 or 0 | Two radii | Inner and outer spheres | $\Phi_0, \Phi_4$ | $2s^2$ |
| Relators | 1 | 3 | Two points | Two poles | $\Phi_1, \Phi_5$ | $2p_x^2$ |
| Metors | 2 | 2 | Two lines | Two semicircular arcs | $\Phi_2, \Phi_6$ | $2p_y^2$ |
| Vitors | 3 | 1 | Two planes | Two hemispherical surfaces | $\Phi_3, \Phi_7$ | $2p_z^2$ |

Time and space are not necessarily symmetry in each quadrant, but when we consider four quadrants together, time and space are indeed symmetry. For example, the relator is in symmetry to the vitor with regard to space and time switch. The scalar of the inner sphere is in symmetry to the scalar of the outer sphere. And the metor, bearing two dimensions of space and two dimensions of time, is spacetime symmetry by itself.

Within a spherical layer, each of the octet (the scalars, the vitors, the metors, and the relators) contains four dimensions of space or time. Space and time dimensions are complimentary so that the total number of dimensions in any quadrant remains four. For a specific quadrant, if its space dimension reduces, then its time dimension must increase accordingly, and vice versa. Space and time components are interwoven (Fig 4). At a specific moment in a dynamic spherical layer, a quartet of scalar, vitor, metor, and relator may experience space expansion with time rarefication while the other counterpart quartet may undergo space reduction with time condensation.

| scalar | relator | metor | vitor | scalar |
|---|---|---|---|---|
| $\Phi$ | $\dfrac{\partial \Phi}{\partial t}$ | $\dfrac{\partial^2 \Phi}{\partial t^2}$ | $\dfrac{\partial^3 \Phi}{\partial t^3}$ | $\dfrac{\partial^4 \Phi}{\partial t^4}$ |
| $\dfrac{\partial^4 \Phi}{\partial l^4}$ | $\dfrac{\partial^3 \Phi}{\partial l^3}$ | $\dfrac{\partial^2 \Phi}{\partial l^2}$ | $\dfrac{\partial \Phi}{\partial l}$ | $\Phi$ |

Fig. 4 The complementary structure of space and time dimensions within four various quadrants.

By the approach of analytic geometry, four distinctive quadrants form four axes of quaternity coordinates (Fig. 5) where every pair of adjacent axes are orthogonal and have a velocity interval. If the space dimension increases, then the time dimension should decrease in the meantime. If we denote the scalar axis by a real number, then the other three axes have spacetime dimensions of $v$, $v^2$, and $v^3$, respectively, where $v$ is a velocity dimension.

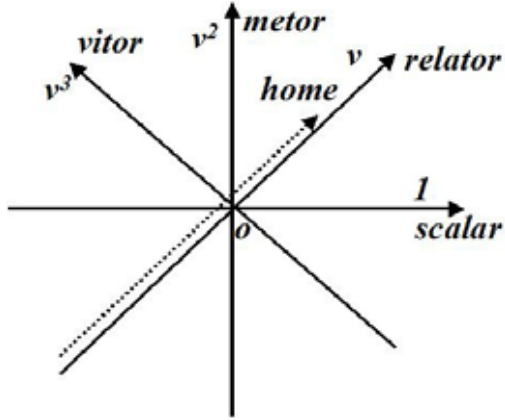

Fig. 5 Illustration of quaternity coordinates with four quadrant axes, where $v$ is a velocity dimension, and the home axis serves as a reference position.

Under quaternity spacetime, we shall consider the transformation of physical quantities between space and time. Given a fixed number of dimensions for a physical entity, reducing a dimension in time is accompanied by increasing a dimension in space, i.e., a differentiation upon a quadrant with respect to a time dimension must be accompanied by an integral operation over a space dimension, and vice versa. For this reason, we shall invent rotation operation in quaternity spacetime as follows. In quaternity coordinates, a counterclockwise rotation of the home axis reduces a time dimension and increases a space dimension. If $\Phi$ is a spacetime function, then its rotation $\Phi^0$ is defined as:

$$\Phi^0 \equiv \int \left(-\frac{\partial \Phi}{\partial t}\right) dl \qquad (20)$$

or

$$\frac{\partial \Phi^0}{\partial l} = -\frac{\partial \Phi}{\partial t} \qquad (21)$$

where $l$ represents the highest time dimension within $\Phi^0$, $t$ represents the highest time dimension within $\Phi$, and the minus sign indicates that space and time dimensions are anti-parallel, varying in opposite directions. Letter $l$ is a general expression of $l_0$, $l_1$, $l_2$, or $l_3$ as were described above, and dimension $t$ has the same implication. For example, if the current home axis is at the location of three-dimensional time with one-dimensional space, then a rotation operation brings the relator to the location of two-dimensional time with two-dimensional space where $l$ denotes $l_2$ in the resultant metor and $t$ indicates three-dimensional time element of the relator. If the current home axis is at the metor, then a rotation operation brings it to the vitor position.

Furthermore, as was demonstrated in eq. (16), electronic transformation follows rotation operation. It can be seen from Fig. 5 that rotating the home axis eight times consecutively brings it to the same axis position; and rotating it four times consecutively reaches the same axis but in the opposite direction. Since neon shell is an inert conservative system, a proper wavefunction describing electronic oscillation within neon shell must satisfy the following equations:

$$\Phi^{08} = \Phi \tag{22}$$

$$\Phi^{04} = C_3 \Phi \tag{23}$$

where $\Phi^{08}$ and $\Phi^{04}$ denote rotating counterclockwise eight times and four times from $\Phi$ respectively, and $C_3$ is a proportionality constant. Since the distance between $\Phi^{04}$ and $\Phi$ is four axes apart, $C_3$ actually bears a dimension of $v^4$. It is very important to realize that both equations hold true no matter where its original home axis is. For example, if the home axis is at the four-dimensional time position, then the equations are for the time function $\Phi$. Otherwise, if the home axis is at the three-dimensional time with one-dimensional space location, then the equations are for the spacetime function $\Phi$. Eq. (23) can be expressed in partial derivative form as:

$$\frac{\partial^4 \Phi}{\partial t^4} = v^4 \frac{\partial^4 \Phi}{\partial l^4} \tag{24}$$

which is called quaternity equation. The solution to it will give wavefunctions of eight electrons in neon shell. Even though the eight electronic orbitals have four different kinds of space and time combinations, each of the scalars, the vitors, the metors, and the relators obeys eq. (24) individually and simultaneously.

## 5. Electronic orbitals in neon shell

Since electrons in neon shell occupy a spherical layer outside helium shell, in order to derive wavefunctions for 2s2p electrons, we build an initial function upon the handler of $1s^2$ electrons:

$$\Phi_{01} = C_4 \cos\Omega_0 \cos\Omega_2 \tag{25}$$

where $C_4$ is a vibration amplitude, and functions $\Omega_0$ and $\Omega_2$ represent time and space radian angles respectively. The rationale for constructing this initial function will be explained in section 6. The first four wavefunctions $2s^1 2p_x^1 p_y^1 p_z^1$ are obtained by performing consecutive rotation operations on $\Phi_{01}$:

$$\begin{pmatrix} \Phi_0 \\ \Phi_1 \\ \Phi_2 \\ \Phi_3 \end{pmatrix} = \begin{pmatrix} \Phi_{01} - \dfrac{\partial \Phi_{01}}{\partial t} \\ v(\Phi_{01}^0 - \dfrac{\partial \Phi_{01}^0}{\partial t}) \\ v^2(\Phi_{01}^{02} - \dfrac{\partial \Phi_{01}^{02}}{\partial t}) \\ v^3(\Phi_{01}^{03} - \dfrac{\partial \Phi_{01}^{03}}{\partial t}) \end{pmatrix} \tag{26}$$

where $\Phi_{01}^{02}$ and $\Phi_{01}^{03}$ denote rotation operations upon $\Phi_{01}$ twice and thrice. Substituting the value of $\Phi_{01}$ into eq. (26) while exploiting the relationship of (16) produces

$$\begin{pmatrix} \Phi_0 \\ \Phi_1 \\ \Phi_2 \\ \Phi_3 \end{pmatrix} = C_4 \begin{pmatrix} \cos\Omega_0 \cos\Omega_2 - \dot{\Omega}_0 \sin\Omega_0 \cos\Omega_2 \\ -v(\sin\Omega_0 \sin\Omega_2 + \dot{\Omega}_0 \cos\Omega_0 \sin\Omega_2) \\ v^2(\cos\Omega_0 \cos\Omega_2 - \dot{\Omega}_0 \sin\Omega_0 \cos\Omega_2) \\ -v^3(\sin\Omega_0 \sin\Omega_2 + \dot{\Omega}_0 \cos\Omega_0 \sin\Omega_2) \end{pmatrix} \quad (27)$$

$$\dot{\Omega}_0 = -\frac{\partial \Omega_0}{\partial t} \quad (28)$$

These four wavefunctions form four quadrants, namely scalar, relator, metor, and vitor respectively in Tab. 1, the other quartet being their counterparts in the opposite direction of quaternity axes that express reverse spacetime dimensions of electronic orbitals within neon shell (Fig. 6).

$$\Phi_{41} = C_4 v^4 \cos\Omega_0 \cos\Omega_2 \quad (29)$$

$$\begin{pmatrix} \Phi_4 \\ \Phi_5 \\ \Phi_6 \\ \Phi_7 \end{pmatrix} = \begin{pmatrix} \Phi_{41} + \dfrac{\partial \Phi_{41}}{\partial t} \\ v(\Phi_{41}^0 + \dfrac{\partial \Phi_{41}^0}{\partial t}) \\ v^2(\Phi_{41}^{02} + \dfrac{\partial \Phi_{41}^{02}}{\partial t}) \\ v^3(\Phi_{41}^{03} + \dfrac{\partial \Phi_{41}^{03}}{\partial t}) \end{pmatrix} \quad (30)$$

which expands into trigonometric expressions as

$$\begin{pmatrix} \Phi_4 \\ \Phi_5 \\ \Phi_6 \\ \Phi_7 \end{pmatrix} = C_4 v^4 \begin{pmatrix} \cos\Omega_0 \cos\Omega_2 + \dot{\Omega}_0 \sin\Omega_0 \cos\Omega_2 \\ -v(\sin\Omega_0 \sin\Omega_2 - \dot{\Omega}_0 \cos\Omega_0 \sin\Omega_2) \\ v^2(\cos\Omega_0 \cos\Omega_2 + \dot{\Omega}_0 \sin\Omega_0 \cos\Omega_2) \\ -v^3(\sin\Omega_0 \sin\Omega_2 - \dot{\Omega}_0 \cos\Omega_0 \sin\Omega_2) \end{pmatrix} \quad (31)$$

These wavefunctions are all constructed via rotation operation from the function of $\Phi_{41}$, which is in turn derived from $\Phi_{01}$ by four consecutive rotation operations. Scalar $\Phi_0$ refers to a four-dimensional time with zero-dimensional space quantity whereas $\Phi_4$ is a four-dimensional space with zero-dimensional time sphere. Both quantities are at the same quaternity axis but in the opposite directions, which are distinguished by a factor $v^4$.

Physically, being in the opposite direction of a quaternity axis has two effects: i) the derivative terms in eq. (30) are positive instead of negative in eq. (26) in the rotation operation pathway; and ii) rotation operation has a reverse meaning, i.e., reducing a time dimension actually means reducing a space dimension, and increasing a time dimension means increasing a space dimension.

$$v^4 \cdot v^i = v^{4-i}, i = 1,2,3,4 \quad (32)$$

This seemingly contradictory arithmetic can be understood in a simple way. For example, as you travel towards south, after passing through the Antarctic pole, you are actually heading towards north if you still maintain the original traveling direction of "south". Similarly, wavefunctions exceeding the extremum of $v^4$ begins to wrap back. In fact, we have seen in eqs. (14) and (15) that spacetime wrap back in the two-dimensional case, i.e., reducing a time dimension from $\Omega_2$ actually has the effect of increasing a time dimension. The first effect is really a corollary to the second. Because in the opposite direction of a quaternity axis, the derivative with respect to time actually means the derivative with respect to space, we have to switch the minus sign in the former into positive to satisfy the contrary tropisms of space and time dimensions.

Wavefunctions in eqs. (27) and (31) form four pairs of electrons with opposite revolving directions, each pair being located at the same quaternity axis but in the opposite directions (Fig. 6). Wavefunctions $\Phi_0$ and $\Phi_4$ are symmetry about space and time and constitute a pair of conjugated complex numbers. Wavefunction $\Phi_5$ matches up with $\Phi_1$ as a pair of relators whereas $\Phi_6$ pairs with $\Phi_2$ as metors, and $\Phi_7$ pair with $\Phi_3$ as vitors. Each pair is a pair of conjugated complex functions. This will become lucid as we further demystify the complex number notations in the next section.

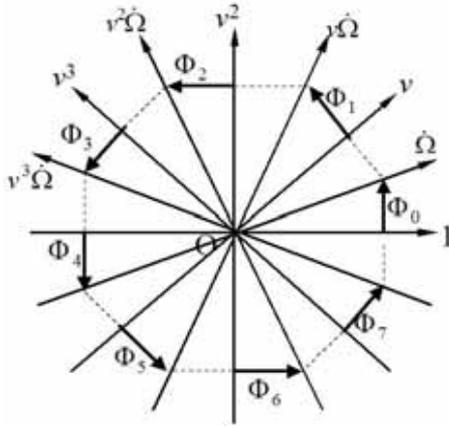

Fig. 6 Dimensional relations between four conjugated pairs of electrons in the opposite sides of quaternity coordinates where every two adjacent axes have a time or space dimension distance.

Among eight wavefunctions, there are invariably rotation relations between adjacent ones, e.g. between $\Phi_0$ and $\Phi_1$, $\Phi_1$ and $\Phi_2$, $\Phi_7$ and $\Phi_0$, and so on. Within neon shell, $2s^2 2p^6$ electrons circulate according to the principle of rotation operation. Electron $\Phi_0$ is traveling in quaternity spacetime and evolving into $\Phi_1$ while $\Phi_1$ is changing into $\Phi_2$, and $\Phi_2$ into $\Phi_3$, and so on. Each electron is gradually shifting its physical state in the following link direction:

$$\Phi_0 \mapsto \Phi_1 \mapsto \Phi_2 \mapsto \Phi_3 \mapsto \Phi_4 \mapsto \Phi_5 \mapsto \Phi_6 \mapsto \Phi_7 \mapsto \Phi_0 \tag{33}$$

After going through eight electronic states, each electron returns to its original waveforms, and another cycle begins. Electronic circulation constitutes harmonic oscillations prettily expressed by quaternity equation. Eight electrons fulfill a complete cycle in rotation transformation at any moment, which explains satisfactorily why a full octet configuration in an atom is a stable system.

From the standpoint of a specific electron, quaternity spacetime includes not only the current electron itself but also the other seven electrons in the past and future of the current electron, and the past and future states are interconnected forming the whole neon sphere. Eight electrons represent eight spacetime dimensions, which give dimensional graduations in spacetime, comprising spatial sphere of the history and future of a specific electron. The neon spacetime is a harmonic oscillation cycle featured by physical electrons. This cyclic spacetime worldview is more fundamental than linear Euclidean X, Y, and Z with infinitely long axes.

## 6. Dimension curling up mechanism

This section is the most abstract and difficult part of the entire paper. As we derived wavefunctions for electrons in neon shell, we simply treated 1s wavefunctions $\Omega_0$ and $\Omega_2$ as two radian angles in eq. (25). What is the basis for this usage? And what is the significance of trigonometric functions? First of all, we shall clarify the relationship between trigonometry and calculus. The relation between differential operation and trigonometric function can be expressed by

$$-\frac{\partial \Omega_0}{\partial t} = C_5 \cos\alpha, where\ (\alpha \mapsto \frac{\pi}{2} + \alpha) \qquad (34)$$

where $C_5$ is a constant. As was mentioned in section 2, differential operation is realized through the revolution of $\alpha$ angle from $\alpha$ to ($\pi/2 + \alpha$), which of course results in the change of $\cos\alpha$ term into $-\sin\alpha$ in the end. Thus, we may express a differential operation in term of radian angle change. Since electronic motion is a dynamic process, we interpret the differential operation as a gradual course. Treating $\alpha$ as a continuously changing variable, we may omit the subordinate clause in eq. (34) and express a differential operation by a trigonometric function, disregarding whether the differential process is carried out completely or not. If it is completely done, then $\alpha$ increases the amount of $\pi/2$ so that $\cos\alpha$ transforms into $-\sin\alpha$. Traditionally, we associate the differentiation result of eq. (34) with the later term, and the initial condition with the former term. Since $\alpha$ is a dynamic variable, the difference of both trigonometric terms is a matter of $\alpha$ value variation within $\cos\alpha$. Both $\cos\alpha$ and $-\sin\alpha$ can be used for expressing a differential operation.

We have seen in eqs. (27) and (31) that trigonometric functions may express wavefunctions in more details than calculus formulae (26) and (30) in a spherical layer. We shall further explain this in situation connecting both 1s and 2s2p spherical layers. Looking at the left-hand side of eq. (34), we may interpret function $\cos\alpha$ as a dynamic fraction where the denominator is the magnitude of a full time dimension $t$ and the numerator is the current time quantity $\Omega_0$. The partial differentiation notation indicates the reduction process of $\Omega_0$ from a full dimension to vanishing, which corresponds to $\alpha$ angle revolution from 0 to $\pi/2$. Thus, logically, when expressing a spacetime component changing from a full dimension to vanishing, we may adopt cosine or sine operator upon a proper radian angle. For the first term of 1s wavefunction in helium shell, we write it as $C_5 \cos\alpha$. But how about the first term of 2s orbital in neon shell? We cannot use $\alpha$ any more because it is a radian angle measuring the internal exchange of space and time components between two 1s electrons cyclically. To express a spacetime extension from 1s to 2s, we write the first term of 2s electron as component $\Phi_{01}$, representing the result of a rotation operation from 1s electron. For two-dimensional quantity $\Omega_0$ in a neon atom, after performing rotation operation once, we have exhausted the time component contained within it. Therefore further differentiation with respect to time must be performed on the whole quantity itself and expressed as $\cos\Omega_0$, i.e., we regard whole $\Omega_0$ quantity as a time radian in neon shell. By the same manner, we have

$$-\frac{\partial \Phi_{01}}{\partial t} = C_6 \cos\Omega_0, where\ (\Omega_0 \mapsto \frac{\pi}{2} + \Omega_0) \qquad (35)$$

which parallels eq. (34) for the subsequent time dimension reduction.

Similarly, $\cos\Omega_2$ is the expression of second integration over a space dimension after extracting the space component within $\Omega_2$. Here we view $\Omega_0$ and $\Omega_2$ as pointers pointing toward the traveling head of both 1s electrons respectively. They are curved vectors or arcs that may represent angles, so the casts of trigonometric operator on $\Omega_0$ and $\Omega_2$ are justified mathematically. We treat $\Omega_0$ as a time radian and $\Omega_2$ as a space radian in neon shell even though each of them is a curvilinear vector containing space and time components in helium shell.

Functions $\cos\Omega_0$ and $\cos\Omega_2$ in neon shell measure significant time and space intervals. For a given length of arc $\Omega_2$, the magnitude of $\cos\Omega_2$ reflects its curvature and hence is an indicator of the curved circular diameter. To specify, when the arc is in the range of (0, π/2), the more severely it curves, the smaller its curved circular diameter, and the smaller the $\cos\Omega_2$ value. Since $\cos\Omega_0$ and $\cos\Omega_2$ are two orthogonal diametrical indicators of a coiled time and space structure, the product of them represents the spacetime block enclosed by two 1s electrons. As we built function $\Phi_{01}$ based on $\cos\Omega_0 \cos\Omega_2$, we were actually dealing with a larger spatial sphere, i.e., using helium shell as a unit to construct neon shell. In other words, we treated $\cos\Omega_0 \cos\Omega_2$ as zero-dimensional space with four-dimensional time in the scope of neon shell and derived wavefunctions for 2s2p electrons according to rotation rule beginning from that point. Our wavefunction construction confirmed the phenomena of dimensions curling up and instantiated the mechanism for curling up a sphere or two dimensions of helium shell as we defined quaternity spacetime in neon shell (Fig. 7).

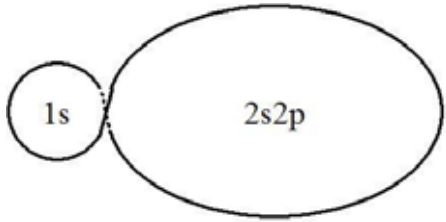

Fig. 7 The two dimensions of 1s electrons are curled up or hidden) in the scope of neon shell where electrons seem to follow the apparent equation (33) path.

Finally, in eqs. (27) and (31), $\dot{\Omega}_0$ communicates with inner $1s^2$ electrons because it may propagate through eqs. (12) to (15). On the other end, $\dot{\Omega}_0$ is generated by a 2s waveform of $\cos\Omega_0$ in a differential chain rule:

$$\frac{\partial(\cos\Omega_0)}{\partial t} = -\sin\Omega_0 \frac{\partial \Omega_0}{\partial t} = \dot{\Omega}_0 \sin\Omega_0 \tag{36}$$

where $\dot{\Omega}_0$ actually acts as a gateway between helium and neon shells. The copulation of both shells follows differential chain rule mathematically. Such an elegant relationship between both 1s and 2s layers confirms the rationale of the above trigonometric curling up mechanism.

The gateway function is also a complex number identifier like $i$. This becomes self-evident when we compare the wavefunction of $\Phi_0$ with that of a 1s electron having a form of

$$\Omega_0 + \Omega_1 = C_1 C_2 (\cos\alpha \cos\beta - \omega \sin\alpha \cos\beta) \tag{37}$$

where $\omega$ is another expression of $i$ as was indicated by eq. (10). The relationship between $\dot{\Omega}_0$ and $\cos\Omega_0$ in eqs. (27) and (31) is similar to that between $\omega$ and $\cos\alpha$ in eq. (37). As was mentioned in section 2, the magnitude of $\omega$ determines the speed of $\alpha$ rotation, so does the magnitude of $\dot{\Omega}_0$ determine the speed of $\Omega_0$ rotation in $\cos\Omega_0$. This means that the inner 1s electrons drive the motion of the outer 2s2p electrons. They are tightly coupled together through $\Omega_0$ and $\Omega_2$, which are two wavefunction pointers to 1s electrons and serve as two radian angles in 2s2p shell. Thus, the derivative term $\dot{\Omega}_0$ is indeed a dimension indicator, a complex number notation, and a gateway function. The wavefunctions of 1s and 2s2p electrons are spacetime continuous. They are in a uniform that features sine and cosine functions, complex function expressions, and gateway communication that synchronizes waves in both layers. In these ways, we have plotted a unique version of multi-dimensional hyperspace where ten dimensions as were represented by 1s2s2p orbitals are contiguous via ten consecutive rotation operations.

## 7. Quaternity equation versus Schrödinger's equation

Quaternity description of electronic orbitals is not only self-consistent but also compatible with well-established quantum physics. This section discusses the junction of quaternity equations and Schrödinger's equations and their divergence therefrom. We shall derive a common heat equation from both two-dimensional spacetime concept in helium shell and Schrödinger's one-orientational equation.

On quaternity side, because space and time components are relative and symmetry in helium shell, the derivative of a wavefunction $\Omega$ with respect to time and its derivative with respect to space must be equal:

$$-\frac{\partial \Omega}{\partial t} = v \frac{\partial \Omega}{\partial l} \tag{38}$$

where the negative sign is introduced because space and time are anti-parallel, and $v$ denotes the dimensional difference. Furthermore, since two electrons within a helium atom are converting between each other, the quantity of one electron is proportional to the rates of changes in another electron and in itself. Let $\Omega_0$ and $\Omega_2$ be the two electrons. We have a quantitative dynamic budget equilibrium relation:

$$\Omega_0 = C_7 \frac{\partial \Omega_0}{\partial t} - C_8 \frac{\partial \Omega_2}{\partial t} + C_9 \tag{39}$$

where $C_i$ are constant parameters. Since space and time are symmetry, when one electron wholly occupies space $\Omega_0$, another electron $\Omega_2$ should be and exactly full time component as was mentioned in section 2. Appling this special boundary condition to eq. (39), we get zero for the first and the third terms so that we have

$$\Omega_0 = -C_8 \frac{\partial \Omega_2}{\partial t} \tag{40}$$

in its simplest form. Since at this moment, space and time, as represented by $\Omega_0$ and $\Omega_2$, respectively, are symmetry and must have a similar shape in mathematic expression, we therefore rewrite eq. (40) as:

$$\Omega = -\frac{1}{\omega} \frac{\partial \Omega}{\partial t} \tag{41}$$

where $\omega$ is a dimension compensator denoting a reciprocal time dimension. Comparing eqs. (38) and (41), we also get

$$\Omega = r\frac{\partial \Omega}{\partial l} \tag{42}$$

Substituting $\Omega$ value of this equation into the right-hand side of eq. (38) produces

$$-\frac{1}{\omega}\frac{\partial \Omega}{\partial t} = r^2 \frac{\partial^2 \Omega}{\partial l^2} \tag{43}$$

This equation is the shape of a well-known thermal diffusion equation or heat equation. It is also called Fick's second law when applied to characterize concentration or molecular diffusion, the diffusion coefficient being assigned to $-\omega r^2$ in this case. Thus it is also proper to say that electronic motion follows the diffusion law.

On quantum mechanics side, Schrödinger's equations for the motion of a particle are the starting point for the development of quantum theory. For a free electron, one-dimensional differential equation is as follows.

$$i\hbar \frac{\partial \phi}{\partial t} = -\frac{\hbar^2}{2m}\frac{\partial^2 \phi}{\partial x^2} \tag{44}$$

In order to compare this equation with eq. (43), we write down the following basic physical relationships:

$$r = \frac{\lambda}{2\pi} \; ; \; \omega = 2\pi f \tag{45}$$

$$E = hf \; ; \; p = \frac{h}{\lambda} \tag{46}$$

$$\hbar = \frac{h}{2\pi} \tag{47}$$

$$E = \frac{p^2}{2m} \tag{48}$$

where $h$, $\lambda$, $f$, $E$, and $p$ refer to Planck's constant, wavelength, frequency, energy, and momentum, respectively. When studying the kinematics of an oscillating object, parameters $\omega$ and $r$ are more descriptive and pertinent than $\hbar$ and $m$. They are closely related to energy and momentum through the rationalized Planck's constant:

$$\omega = \frac{E}{\hbar} \; ; \; \frac{1}{r} = \frac{p}{\hbar} \tag{49}$$

After converting parameters $\hbar$ and $m$ into $\omega$ and $r$, eq. (44) becomes

$$-\frac{1}{\omega}\frac{\partial \phi}{\partial t} = r^2 \frac{\partial^2 \phi}{i\partial x^2} \tag{50}$$

Under quaternity spacetime, we interpret the denominator $i\partial x^2$ as the generalized space dimensions $\partial l^2$ so that eqs. (43) and (50) are equivalent. The two space dimensions contained in $\partial l^2$ in eq. (43) are orthogonal and have different meanings under Euclidean geometry. We may use $\partial x$ to indicate the first dimension of $\partial l$ within wavefunction $\phi$ and use $i\partial x$ to denote the successive space dimension $\partial l$ within $\phi$. In this sense, the complex number identifier represents the shifting of the space dimension order, and has the effect of rotating a space dimension to its perpendicular orientation under Euclidean geometry. This is in consistent with our original interpretation of complex number identifier $i$ or $j$ where it transforms its operand to the dimension orthogonal to it. Thus, the one-dimensional heat equation is the common ground of quaternity equations and Schrödinger's equations. After all, they diverge into different paths thereafter due to their different perceptions on spacetime.

Quantum mechanics extends one-dimensional Schrödinger's equation into three-dimensional by introducing Laplacian operator:

$$i\hbar \frac{\partial \phi}{\partial t} = -\frac{\hbar^2}{2m} \nabla^2 \phi \qquad (51)$$

which sees three dimensions of space in X, Y, and Z orientations. By using Laplacian operator $\nabla^2$, it is implied that electronic orbitals distribute equally in X, Y, and Z directions, i.e., that space is homogenous and isotropic. Notwithstanding this impression, quantum mechanics traditionally handled Schrödinger's equation by transforming Cartesian coordinates into spherical polar coordinates pertinently and then tried to derive quantum information under the constrains that the equation must have solutions and that the solutions must be converged. Only via this transformation, it is successful in getting much useful information on electronic orbitals.

In contrast, quaternity views the spaces represented by X, Y, and Z axes as possessing different levels of dimensions. As shown in Fig. 8, supposing that we are trying to locate a particular seat at a great auditorium, we would first look for the floor number as was represented by Z direction, and then within the proper floor search the tier number as was represented by Y axis, and finally locate the X position in that tier. This simple logic is not beyond our everyday living experience, but mathematicians so far have not caught the significant difference between the three axes. Here, three-dimensional Z direction refers to the arrangement of all the floors, each of which contains multiple tiers; two-dimensional Y direction refers to the sequence of the tiers, each of which contains multiple seats; and one-dimensional X direction refers to the order of seats. The wavefunctions for the three orientations represent different meanings and constitute three orthogonal dimensions. The natural shifting of the dimensions from floor order, to tier order, to seat order in locating a particular seat at the auditorium reflects the rule of quaternity differentiation, which is always operated with respect to the most immediate dimension. In formulating multi-dimensional wave equations, quaternity climbs up to higher rank of derivatives and sees four space dimensions in $\partial/\partial l$, $\partial^2/\partial l^2$, $\partial^3/\partial l^3$, and $\partial^4/\partial l^4$ forms instead of adopting Laplacian operator (also see eq. 24 and Fig. 4).

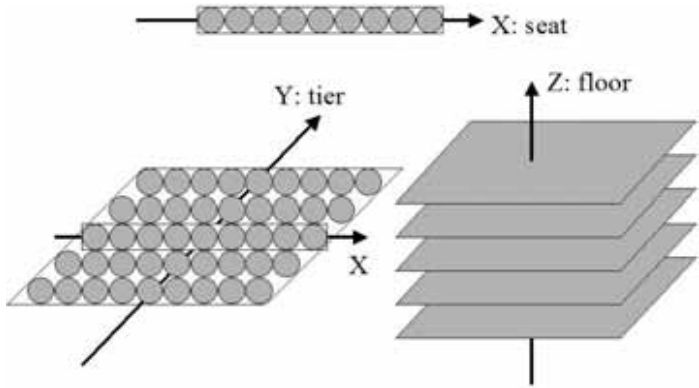

Fig. 8 Even in a theater ticket, the meanings of X, Y, and Z axes in Euclidean geometry denote different levels of space dimensions.

Since the second derivative contains the process of the first derivative, wave eq. (43) is actually two-dimensional in helium spacetime. Moreover, because the fourth derivative term contains the operations of the third, second, and first derivatives, quaternity equation (24) governs electronic oscillations in four various dimensions indeed. As was demonstrated, the solution to this wave equation gave eight electronic wavefunctions in neon shell. Electrons exist in various spacetime niches. In neon shell, eq. (24) means that performing rotation operations on an electron four times consecutively returns electronic state to the same quaternity axis but in the opposite direction; and performing rotation operations eight times sequentially produces eight different electronic states that form 2s2p octet.

## 8. Prediction and evidences of 2p-orbital geometries

After describing the geometrical motion of 2s2p electrons, we have more confident to predict and retrodict the geometrical shapes of 2p-oribtals. Our description of $2p_x$, $2p_y$, and $2p_z$-orbitals indicate that each orbital type is different geometrically. Assuming that electron cloud distributes in X, Y, and Z orientations equally and statically cannot explain the orientation of 2p-orbitals in carbon atoms that form methane, ethylene, and ethyne molecules. Hypothetical orbital hybridizations are traditionally supplemented for those accounts.

Quaternity spacetime describes 2p-orbitals using two relators, two metors, and two vitors that have different geometric shapes in spherical polar coordinates ($\rho$, $\theta$, $\varphi$). Two 2s orbitals are scalars of spherical shape defined by $0<\rho<R$ where R is the spatial radius of a neon shell. As shown in Fig. 9, a $2p_x$ is a relator of a linear vector in either pole constrained by $0<\rho<R$ and ($\theta=0$ or $\pi$) and $\varphi=\pi/2$ in spherical polar coordinates; a $2p_y$ is a metor of a planar sector bound by $0<\rho<R$ and ($0<\theta<\pi$ or $\pi<\theta<2\pi$) and $\varphi=\pi/2$; and a $2p_z$ is a vitor of hemispherical shape of $0<\rho<R$ and ($0<\varphi<\pi/2$ or $\pi/2<\varphi<\pi$). There are various geometrical flexibilities associated with each orbital. A free electron may also evolve dynamically from 2s to $2p_x$, to $2p_y$, to $2p_z$ in sequence as was illustrated by eq. (33).

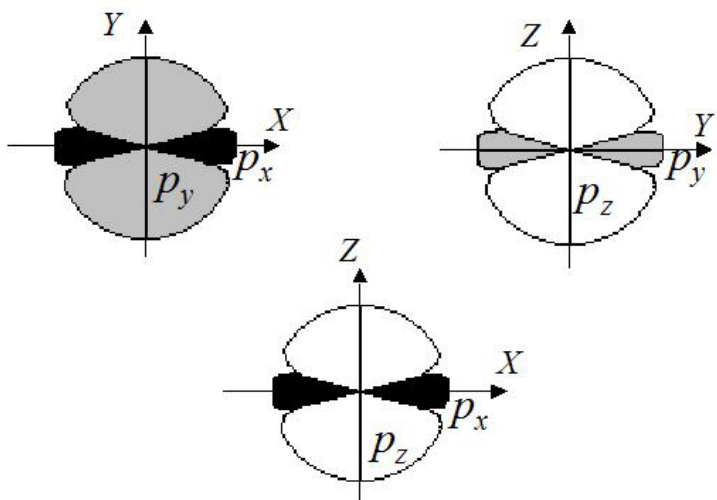

Fig. 9 Profile drawing of geometrical shapes of 2p-orbitals in three-dimensional Euclidean where $2p_x$ are two one-dimensional vectors (black); $2p_y$ are two flat sectors (gray); and $2p_z$ are two hemispheres (white).

The geometrical restrictions and dynamical flexibilities can explain molecule structure of various organic molecules without recourse to orbital hybridization. For example, in a methane molecule, carbon atom uses a 2s electron, a $2p_x$ electron, a $2p_y$ electron, and a $2p_z$ electron to form σ-bonds with four hydrogen atoms. Because of the flexibility of electronic motion, each bond is restricted to a corner of a tetrahedron under electrical repulsions. But the four σ-bonds are all different due to different natures of the four electrons. This explains why four σ-bonds actually have unequal lengths at any moment.

In an ethylene molecule, two carbon atoms form a stable $2p_x$-$2p_x$ σ-bond and a $2p_y$-$2p_y$ π-bond. Geometrically, the π-bond is the partial overlap of two $2p_y$ flat sectors in one atom with two $2p_y$ flat sectors in another atom (Fig. 10). It has less overlap than the σ-bond in the middle and is an unsaturated covalent bond. The geometrical orientation of $2p_y$ electrons to accommodate for π-bond and the geometrical flexibility of 2s and $2p_z$ under electrical repulsion of the σ-bond and π-bond explain its planar molecular conformation.

In an ethyne molecule, carbon atoms form a $2p_x$-$2p_x$ σ-bond, a $2p_y$-$2p_y$ π-bond, and a $2p_z$-$2p_z$ π-bond. The two π-bonds are directional oriented and are of different energy, one being two flat sectors partially overlapping two flat sectors, the other being two hemispheres partially overlapping two hemispheres on the edges. The C-H bonds are oriented towards the opposite sides of the three bonds due to electrical repulsion, giving its linear molecular structure. Thus quaternity description of electrons in a carbon atom naturally accommodates the configuration of 2p-orbitals in $sp^3$, $sp^2$, or sp hybridizations under special circumstances and shows great capacity in explaining the structure of various organic molecules. Orbital hybridization hypothesis is not needed in quaternity spacetime. The exact match of four quadrants with 2s, $2p_x$, $2p_y$, and $2p_z$ orbitals is a basic evidence of quaternity spacetime.

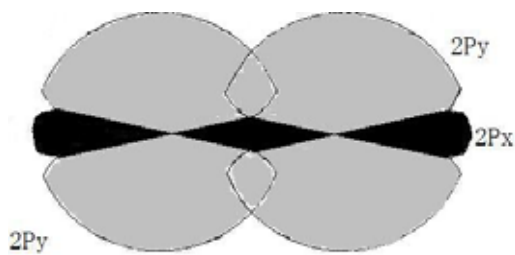

Fig. 10 Schematic diagram of a $2p_x$-$2p_x$ σ-bond (black overlap) and a $2p_y$-$2p_y$ π-bond (gray overlaps) in an ethylene molecule.

In addition, the asymmetrical carbon atom (with four outer electrons of 2s, $2p_x$, $2p_y$, and $2p_z$ structures) explains well with regard to the chirality of molecules in stereochemistry. A chiral molecule is a molecule that is non-superposable on its mirror image. A pair of chiral enantiomers rotate the plane of polarization of plane-polarized light an equal amount, but in opposite directions. We believe that the spacetime differences of 2s, $2p_x$, $2p_y$, and $2p_z$ orbitals account for the light polarizations. Light bends due to spacetime curvatures within the atomic sphere. This phenomenon was observed in chiral molecules only because chiral covalent bonds to an asymmetrical carbon atom prevent each of its four outer electrons from switching among 2s, $2p_x$, $2p_y$, and $2p_z$ orientation states as it normally does dynamically (eq. 33) so that the molecule bends the light in a fixed chirality. The theory of quaternity will undoubtedly provide fresh insight into the configuration of organic molecules and their reaction mechanisms as it penetrates into organic chemistry.

## 9. Summary

Quantum mechanics cannot describe in reasonable details on how electrons move within stable systems such as helium and neon atoms. We believe that this is not the problem of theory construction, but the limitation of its undergirding foundation that we have taken it for granted. This foundation is Euclidean space (or Hilbert space) and Newtonian time. When the basement of space and time concept deviates from the reality, the entire establishment built upon it is flaw. However, since Euclidean space and Newtonian time are deeply ingrained in people's mind and imagination and have dominated the knowledge base for thousands of years, changing them seems out of the question. This poses a daunting challenge for the introduction of quaternity spacetime.

In this paper, we have established quaternity spacetime for describing electronic orbitals in a better precision than quantum mechanics. To specify, we explored space and time concept from scratch with a novel mindset, established four distinctive space dimensions in a spherical layer originally, discovered the mechanism for dimensions curling up for the first time, examined the principle of rotation operation as electronic transformation, and described electronic orbitals with four curvilinear quadrants that oscillate simultaneously in four dimensions. Quaternity spacetime is a revolutionary idea in fundamental geometry and physics.

Nevertheless, quaternity description of electrons is compatible with quantum theory. Quaternity uses four pairs of curvilinear vectors for describing electronic states, which are in correspondence with four quantum numbers. The scalar defines a spherical layer, which corresponds to principal quantum number that accounts for energy level band; the vitor of conical shape corresponds to orbital quantum number that determines the magnitude of orbital angular momentum; the metor in the equatorial plane is associated with magnetic quantum number; and the relator, having one space with three time dimensions, describes intrinsic angular momentum perfectly. These conceptual matches constitute an additional interpretation of quantum numbers. This alternative spacetime theory has been fully expressed in a recent monograph [2].